\documentstyle[psfig,12pt,twocolumn]{article}

\begin{document}

\title{Two Common Misconceptions about the Theory of Special Relativity}
\renewcommand{\baselinestretch}{0.5}

\author{
{\large Ezzat G. Bakhoum}\\
\\
{\small New Jersey Institute of Technology}\\
{\small P.O. Box 305, Marlton, NJ. 08053 USA}\\
{\small Email: bakhoum@modernphysics.org}\\
\\
{\small Copyright \copyright 2004 by Ezzat G. Bakhoum}\\
\\
\\
\begin{minipage}{6in}
\begin{center}
{\bf Abstract}
\end{center}
{\small 
Two common misconceptions about the theory of Special Relativity that are actively taught in textbooks are discussed. It is shown, first, that the Lorentz transformations are actually transformations of the coordinates of a {\em photon}, not the coordinates of a particle as taught by some authors. Secondly, a misconception concerning the relativistic Lagrangian is discussed. It is shown that the currently accepted formulation of the Lagrangian is missing an important constant of integration. By incorporating the missing constant of integration, the new Lagrangian directly supports the conclusions reached previously by the author concerning the mass-energy equivalence principle.}
\end{minipage}
}

\date{}

\maketitle

{\large\bf Preface:}\\
\\
This paper will be the last in my series of papers on the principle of mass-energy equivalence. Since my previous papers have essentially fallen on deaf ears (with the exception of a few, truly remarkable individuals), I have regrettably concluded that it is now time to stop ``preaching to the wind". Not until all the logic-defying excuses that started in 1930 (a.k.a. the Neutrino theories) totally collapse as a result of their ever-increasing state of chaos, can someone who is truly intelligent realize that something is potentially wrong with the current state of subatomic physics.\\
\\
\\
{\large\bf 1. Misconception concerning the Lorentz Transformations:}\\
\\
It is regrettable that the Lorentz transformations are introduced in many of the contemporary references on Special Relativity (SR) \cite{French, Jackson} as the transformations of the coordinates of ``an object in motion", such as a particle. A careful examination of Einstein's argument in his paper of 1905 \cite{Einstein}, however, leaves no doubt that the Lorentz transformations are indeed transformations that describe the coordinates of a {\em photon}. The error of assuming that the transformations describe the coordinates of a material object is usually made because of the failure to distinguish between the coordinates of an object and the coordinates of an ``event" in space-time that is described, according to SR, by the propagation of a light signal. This observation, of course, will have very important implications for the various applications of the Lorentz transformations, and, in particular, the principle of mass-energy equivalence.\\
\\
According to the 1905 paper, the following is the argument of SR: if a photon is propagating and is observed simultaneously in two frames of reference $S$ and $S^\prime$ (where $S$ is the stationary frame and $S^\prime$ is the moving frame and where the propagation starts when the origins of the two frames coincide), then $x = ct$ and $x^\prime = c t^\prime$ are two experimentally verified relationships that we must hold as true. Hence, in order to accommodate the law of the constancy of the velocity of light in the two frames, $t \neq t^\prime$ (time must be somehow distorted by a factor $\gamma$). Accordingly, any Newtonian distance must also appear to be distorted by the same factor $\gamma$. The distance traversed by the moving frame $S^\prime$, being equal to $vt$, will be therefore given by $v(\gamma t^\prime) = \gamma v t^\prime$. Hence the $x$ coordinate of the {\em photon} will be 

\begin{eqnarray}
x & = & \gamma x^\prime + \gamma vt^\prime \nonumber\\
 & = & \gamma (x^\prime + vt^\prime) 
\label{1}
\end{eqnarray}

This is the first of the Lorentz transformations. The rest of the derivation that leads to the second transformation $t  =  \gamma (t^\prime + vx^\prime / c^2)$ then proceeds by  using the reciprocal relationship and substituting for $x$ and $x^\prime$ by $x = ct$ and $x^\prime = c t^\prime$. It is therefore very important to note that $x$ and $x^\prime$, according to the argument of SR, are strictly the coordinates of a photon, not of a particle.\\
\\
What the coordinates of a particle will be then? Since the particle is assumed to be {\em at rest} in the moving frame $S^\prime$, then $x^\prime = 0$, and hence the Lorentz transformations take the reduced form

\begin{eqnarray}
x & = & \gamma vt^\prime \nonumber\\
t & = & \gamma t^\prime 
\label{2}
\end{eqnarray}

The conclusion therefore is that the full Lorentz transformations apply {\em only} to the coordinates of a propagating photon. For a material body that is assumed to be at rest in the moving frame $S^\prime$, the reduced Lorentz transformations of Eq.(\ref{2}) will be the correct transformations to apply. Using the full transformations to describe the coordinates of a particle does indeed lead to serious errors and inexplicable results. As was pointed out in \cite{Bakhoum2}, de Broglie's use of the full Lorentz transformation of time is essentially what led him to the amazing velocity $c^2/v$. However, it was also pointed out in \cite{Bakhoum2} that the full transformations can still be used to transform quantities such as velocity and momentum of a material particle, provided that we pay special attention to the constant $c$ that is attached to the time coordinate. As was shown in that paper, the full transformations can be successfully used only if we treat $c$ as a ``light like" velocity component that has a non-vanishing derivative with respect to time.\\
\\
While it is true that the Lorentz transformations represent a ``mapping" of a vector from one space into another, we must realize that this vector is strictly defined by SR as the vector describing the coordinates of a propagating photon (the definition of an ``event" in space-time). The argument of SR was not based at all on the kinematics of subluminal objects. Therefore special care and attention must be exercised when incorporating the kinematics of subluminal objects into SR. This was shown in ref.\cite{Bakhoum2}. But while it is also absolutely true that the kinematics of light propagation that SR describes leads to what seems to be a distortion in space and time, the reduced transformations of Eq.(\ref{2}) show that by setting $x^\prime = 0$, this important notion (that is, the relativity of space and time) becomes {\em coordinate independent}; in other words, the moving particle itself becomes the reference frame $S^\prime$.  The coordinate {\em dependent} transformations, i.e., the full transformations, on the other hand, represent a more specific level of detail beyond the general concept of the relativity of space and time: a description of the kinematics of light particles.\\
\\
\\
{\large\bf 2. Misconception concerning the Relativistic Lagrangian:}\\
\\
{\bf 2.1. The missing constant of integration:}\\
\\
Another common misconception is about the currently accepted formulation of the relativistic Lagrangian. The relativistic Lagrangian was never given by Einstein, Lorentz, or Poincar\'{e}, but was actually derived by a number of modern authors \cite{Jackson, Goldstein}. The derivation of the relativistic Lagrangian typically proceeds as follows: by taking the free-particle equation of motion 

\begin{equation}
F = \frac{dp}{dt} = \frac{d}{dt} (\gamma m_0 v) = 0
\label{3}
\end{equation}

(where $F$ is the force, $p$ is the momentum and $v$ is the velocity), together with the Euler-Lagrange equation of motion

\begin{equation}
\frac{d}{dt} \left(\frac{\partial L}{\partial \dot{q}_i} \right) -
\frac{\partial L}{\partial q_i} = 0
\label{4}
\end{equation}

where $L$ is the Lagrangian and $q_i$ are the coordinates, and given that $\partial L / \partial q_i = 0$ for a free particle, we are then directly led to the condition 

\begin{equation}
\frac{\partial L}{\partial v} = \gamma m_0 v,
\label{5}
\end{equation}

where we have assumed motion along one dimension only for simplicity. Jackson and Goldstein \cite{Jackson, Goldstein} showed that a Lagrangian that will satisfy this equation will be (which can be obtained by direct integration; note that $\gamma = (1-v^2/c^2)^{-1/2}$)

\begin{equation}
L = - m_0 c^2 \sqrt{1-v^2/c^2}
\label{6}
\end{equation}

The problem here obviously is that the possible existence of a constant of integration was {\em completely ignored}. If we integrate Eq.(\ref{5}) between the limits 0 and $v$, 

\begin{equation}
L = \int_0^v \frac{m_0 v}{\sqrt{1-v^2/c^2}} \; dv
\label{7}
\end{equation}

we obtain instead

\begin{equation}
L = m_0 c^2 (1- \sqrt{1-v^2/c^2})
\label{8}
\end{equation}

The missing constant of integration is therefore the quantity $m_0 c^2$ (it is worthwhile to note that Einstein did not ignore the constant of integration in his derivation of the relativistic kinetic energy \cite{Einstein}). Note that for a velocity $v=0$, Eq.(\ref{6}) gives $L= -m_0 c^2$ while Eq.(\ref{8}) gives $L=0$. We should expect $L$ to be equal to 0 at rest, since, in the absence of external forces, the Lagrangian is simply the kinetic energy, and the kinetic energy is equal to 0 at rest. We can indeed confirm that Eq.(\ref{8}) is the correct result by simply noting the following fact about the action integral $A$ \cite{Jackson, Feynman}:

\begin{equation}
A = \int_{t_1}^{t_2} L \; dt
\label{9}
\end{equation}

If we write the action integral in terms of the proper time $\tau$, where $d\tau = dt / \gamma$, it becomes

\begin{equation}
A = \int_{\tau_1}^{\tau_2} \gamma L \; d\tau
\label{10}
\end{equation}

If we now multiply the expression in Eq.(8) by $\gamma$, it becomes

\begin{eqnarray}
\gamma L & = & \frac{m_0 c^2}{\sqrt{1-v^2/c^2}} - m_0 c^2 \nonumber\\
         & = & \Delta m c^2 \nonumber\\
         & = & \mbox{Relativistic Kinetic Energy}
\label{11}
\end{eqnarray}

Hence Eq.(\ref{10}) can be written as

\begin{equation}
A = \int_{\tau_1}^{\tau_2} \left( \mbox{Relativistic Kinetic Energy} \right) \; d\tau
\label{12}
\end{equation}

In the absence of external forces (i.e., potential energy), this is indeed the expression that we should expect to obtain. Perhaps what aided in supporting the misconception is another misconception, namely, that the action integral $A$ in Eq.(\ref{12}) above must be a Lorentz invariant quantity (see the discussion in ref.\cite{Jackson}). However, we must distinguish between Lorentz invariance and {\em path invariance}. The proper time $\tau$ in Eq.(\ref{12}) above is Lorentz invariant but the quantity $A$ is path-invariant (note that this is true only for small variations in the path, due to the extremum condition $\Delta A = 0$). The opposite, however, is not true. $\tau$ is not path-invariant and $A$ should not be automatically assumed to be a Lorentz-invariant quantity.\\
\\
{\bf 2.2. The relationship between the Lagrangian and the Hamiltonian:}\\
\\
If we now take the corrected expression of the Lagrangian in Eq.(\ref{8}) and let $v \approx c$, we obtain $L = m_0 c^2$. This result suggests that there is, at least, a relationship between the Lagrangian and the mass-energy quantity (it was demonstrated in ref.\cite{Bakhoum1} that the quantity $ m_0 c^2 (1- \sqrt{1-v^2/c^2})$ is indeed the mass-energy of a particle). To put things into perspective, we now make the following distinction between the definition of the Lagrangian in classical mechanics and its definition in relativistic mechanics: in the classical (Newtonian) mechanics, the Lagrangian of a free particle (that is, in the absence of external forces) is simply its kinetic energy. In relativistic mechanics, Eq.(\ref{11}) shows that the Lagrangian is given by the ratio of the relativistic kinetic energy to the Lorentz gamma factor, or $E_k / \gamma$. Now, the emergence of $L = m_0 c^2$ for $v \approx c$ suggests very strongly that the mass-energy of a free particle is equal to the ratio $E_k / \gamma$ (which has an upper limit of $m_0 c^2$). This point was also clear from 
ref.\cite{Bakhoum1}. To summarize, it is quite apparent from these results that

\begin{equation}
L (\mbox{Relativistic}) = \mbox{Mass-Energy} = \frac{E_k}{\gamma}
\label{13}
\end{equation}

What does the above result then mean as far as the concept of the Hamiltonian is concerned? As we know, the Hamiltonian $H$ of a free particle is the sum of the kinetic and the mass energies. Hence,

\begin{eqnarray}
H & = & E_k + \frac{E_k}{\gamma} \nonumber\\
  & = & \left( \gamma m_0 c^2 - m_0 c^2 \right) +
        \left( m_0 c^2 - \frac{m_0 c^2}{\gamma} \right) \nonumber\\
  & = & m_0 c^2 \left( \gamma - \frac{1}{\gamma} \right) \nonumber\\
  & = & \gamma m_0 c^2 \left( 1 - \frac{1}{\gamma^2} \right) \nonumber\\
  & = & \gamma m_0 c^2 \left( 1 - (1 - v^2/c^2) \right) \nonumber\\
  & = & \gamma m_0 v^2 \nonumber\\
  & = & m v^2
\label{14}
\end{eqnarray}

The conclusion therefore is that the (mathematically correct!) Lagrangian given by Eq.(\ref{8}) supports the total energy equation $H = mv^2$. While it is absolutely true that the mathematical machinery provided by SR cannot, by itself, provide a ``derivation" for the mass-energy equivalence principle (as was clearly shown in ref.\cite{Bakhoum1}); it is nonetheless the duty of responsible scientists to derive expressions which are mathematically correct and then make conclusions in favor of one point of view or the other.\\
\\
\\
{\large\bf Acknowledgements:}\\
The author is highly indebted to Prof. Stanley Robertson of the Univ. of Southwestern Oklahoma for many useful discussions about the concepts discussed in this paper.


\begin{thebibliography}{99}

\bibitem{French} A.P. French, {\sl Special Relativity\/} (Norton Publications, New York, NY, 1968).

\bibitem{Jackson} J.D. Jackson, {\sl Classical Electrodynamics\/}
(Wiley, New York, NY, 1975).

\bibitem{Einstein} A. Einstein, {\sl On the Electrodynamics of Moving Bodies\/}
(The Collected Papers of A. Einstein, Vol.2, Princeton Univ. Press, 1989), p.140.

\bibitem{Bakhoum1} E. Bakhoum, {\sl Fundamental Disagreement of Wave Mechanics with Relativity\/}, Physics Essays, 15, 1, 2002, p.87-100. Online e-print archive: physics/0206061.

\bibitem{Bakhoum2} E. Bakhoum, {\sl Dialogue on the Principle of Mass-Energy Equivalence\/}, Physics Essays, 17, 4, 2005 (to appear). Online e-print archive: physics/0402038.

\bibitem{Goldstein} H. Goldstein, {\sl Classical Mechanics\/} (Addison-Wesley, Reading, MA, 1965).

\bibitem{Feynman} R. Feynman et al., {\sl The Feynman Lectures on Physics, Vol.2\/} (Addison Wesley, Reading, Mass., 1964).



\end{thebibliography}
\end{document}